\documentstyle[12pt,epsf]{article}
\def\Journal#1#2#3#4{{#1} {\bf #2}, #3 (#4)}

\def\PLB{{\em Phys. Lett.}  B}

\def\PRD{{\em Phys. Rev.} D}
\def\ZPC{{\em Z. Phys.} C}
\def\PR{\em Phys. Rep.}

\def\be{\begin{equation}}
\def\ee{\end{equation}}
\def\bea{\begin{eqnarray}}
\def\eea{\end{eqnarray}}

\begin{document}
\def\singlespacing{\baselineskip=12pt}
\def\doublespacing{\baselineskip=20pt}
\doublespacing
\pagestyle{empty}

\noindent
February 2nd, 1996 \hfill {\bf MC-TH-96/06}

\vspace{2cm}

\begin{center} 
\begin{large}

{\bf BOUND STATES OF TWO GLUINOS AT THE TEVATRON AND LHC}  

\bigskip

{\em E. Chikovani\/${}^{a}$, V. Kartvelishvili\/${}^{a,b}$, \\
R. Shanidze\/${}^{a}$ and G. Shaw\/${}^{b}$}

\end{large}

\bigskip

${}^{a}$~High Energy Physics Institute, Tbilisi State University,\\
Tbilisi, GE-380086, Republic of Georgia.

\bigskip

${}^{b}$~Department of Physics and Astronomy, Schuster Laboratory,\\
University of Manchester, Manchester M13 9PL, U.K.

\end{center}

\bigskip

\bigskip

\begin{center}
{\bf ABSTRACT}
\end{center}

\noindent
We calculate the production cross-sections for the vector and pseudoscalar
bound states of two gluinos. It is shown that existing and future 
colliders imply a realistic chance of observing  gluinonium
as a narrow peak in the two-jet invariant mass spectrum.
With an integrated luminosity of 0.2 fb$^{-1}$ at the Tevatron, and the high 
efficiency for tagging
heavy quark jets at CDF, one should be able to detect vector gluinonium 
for gluino 
masses up to about 170 GeV; or up to about 260 GeV for an upgraded
Tevatron with a centre of mass energy of 2 TeV and an integrated luminosity
of 1 fb$^{-1}$. The significantly higher energy and luminosity
of LHC should allow pseudoscalar gluinonium to be detected for gluino
masses up to about 1500 GeV for an assumed luminosity of 200 fb$^{-1}$.
These results are
insensitive to the details of supersymmetry models, provided that R-parity
is conserved and the gluinos are lighter than the squarks. In addition,
gluinonium detection implies a  
relatively accurate measure of the gluino mass, which is difficult to 
determine by other means.

\newpage
\pagestyle{plain}
\pagenumbering{arabic}

\section{Introduction}

In most supersymmetry models \cite{Haber},
the ease with which gluinos can be detected depends critically on the relative
masses of gluinos and squarks. If the gluino mass $m_{\tilde{g}}$ 
is greater than the squark mass $m_{\tilde{q}}$, 
the gluino can decay to a squark and antiquark (or an antisquark 
and quark). This is a strong decay  with  a decay width of order $\alpha_s \, 
m_{\tilde{g}}$, which has prominent signatures, so that the
 Particle Data Group \cite{PDG} give a current experimental lower limit
of $m_{\tilde{g}}\geq 218$ GeV. On the other hand, if the squarks are heavier 
than the
gluino and R-parity is conserved, the gluino can only decay  weakly, as shown
in Figure 1, with a width of order \cite{Haber}
\be
\Gamma \approx \frac{\alpha \alpha_S m_{\tilde{g}}}{48 \pi} \left( \frac{
m_{\tilde{g}}}{m_{\tilde{q}}} \right)^4 \approx 5 \times 10^{-6}
\left( \frac{
m_{\tilde{g}}}{m_{\tilde{q}}} \right)^4  m_{\tilde{g}} \; .
\label{i1}
\ee
The gluino is then more difficult to detect and the mass limit is reduced to
 100 GeV \cite{PDG}. However, in the latter case there is the possibility
of observing the gluino indirectly, by detecting a gluino-gluino bound state.
This has the advantage that the conclusions which can be drawn from a search 
for such states hold in a very wide class of supersymmetry models.
In addition, the detection of such a state would lead to a relatively
precise determination of the gluino mass, which could not be obtained easily
by observing the decay products of the gluino itself, as some of these
escape undetected. Here we investigate the
possibility of detecting such states as narrow peaks in the di-jet
invariant mass distributions, in the light of the
significant improvements in heavy quark jet tagging and beam luminosity 
which have been achieved at the Tevatron, and in the light of the high
energies and luminosities expected at the LHC.

\section{Gluinonium}

Gluinos are heavy spin-1/2 fermions which interact strongly via gluon exchange.
Like gluons,  they have no basic electroweak couplings. Hence, in contrast to 
top quarks, they live long enough to form two gluino bound states provided
$ m_{\tilde{g}} < m_{\tilde{q}}$ as indicated above. The resulting spectrum of
states is called gluinonium and has been studied in a number of papers
\cite{Keung}-\cite{Chikovani90}. For heavy gluinos, the low lying
states are essentially Coulombic, with masses $M \approx 2 m_{\tilde{g}}$,
and they have much in common with heavy quarkonia. 
However, there are two important
differences. 

Firstly, the gluinos form a colour octet, so that there are six possible 
colour states for a two gluino system, corresponding to the well-known
decomposition
\be
8\otimes 8 = 1\oplus 8_S\oplus 8_A\oplus 10\oplus \overline{10}\oplus 27\;.
\ee
The single gluon exchange potential $V(r)$ is attractive for the first three,
and is related to the corresponding potential $V_q(r)$ for $q \bar{q}$ pairs
by
\be
V(r) = \left( \frac{9}{4} , \frac{9}{8} , \frac{9}{8} \right) V_{q}
(r)
\ee
for the $(1, 8_S, 8_A)$ states respectively. 

Secondly, the gluino is a
Majorana fermion with just two degrees of freedom. It is its
own antiparticle and the Pauli principle applies to the two gluino system,
thus forbidding certain states.

The resulting spectra of low lying states \cite{Keung}-\cite{Goldman} 
are shown in Table 1. The allowed colour singlet 1 and 
symmetric octet $8_S$ states
have the same $J^P$ values as the charmonium states with $C = 1$; while the
allowed antisymmetric colour octet $8_A$ states
have the same $J^P$ values as the charmonium states with $C = -1$. 
In particular, the lowest lying  colour singlet and symmetric 
colour  octet states  are the  pseudoscalars  
$\eta_{\tilde{g}}^0$ 
and $\eta_{\tilde{g}}^8$ with $J^{P} = 0^{-}$, while
the lowest lying  antisymmetric colour octet state is 
vector gluinonium $\psi_{\tilde{g}}^8$ with 
$J^{P} = 1^{-}$.

The properties of these low lying 
states have been discussed previously \cite{Goldman}-\cite{Chikovani90}.  
Here we simply summarize the main results.  All three $L=0$ states decay 
predominantly via 
gluino-gluino annihilation as shown in Figure 2. Specifically, 
the pseudoscalars 
$\eta_{\tilde{g}}^{0,8}$ decay mainly to two gluons 
\cite{Keung, Kuhn, Goldman} with decay widths 
\be\label{geta0}
\Gamma(\eta_{\tilde{g}}^0 \rightarrow g \, g ) =
\frac{18 \alpha_S^2}{M^2} | R_0(0) |^2 \approx \frac{243}{8} \alpha_S^5 M 
\ee
\be\label{geta8}
\Gamma(\eta_{\tilde{g}}^8 \rightarrow g \, g ) =
\frac{36 \alpha_S^2}{M^2} | R_8(0) |^2 \approx \frac{243}{32} \alpha_S^5 M 
\ee
while  vector gluinonium $\psi_{\tilde{g}}^8$ decays predominantly 
into $q \bar{q}$ pairs \cite{Chikovani89}
with a decay width 
\be\label{gpsi8}
\Gamma(\psi_{\tilde{g}}^8 \rightarrow q \bar{q} ) = N_q
\frac{2 \alpha_S^2}{M^2} | R_8(0) |^2 \approx N_q \frac{27}{64} 
\alpha_S^5 M \; .
\ee
Here  $M \approx 2 m_{\tilde{g}}$ is the gluinonium mass, $N_q$  is the number 
of quark flavours for which the decay is allowed kinematically, 
and  quark masses have been neglected.
$R_{0,8}(0)$ are the  radial wavefunctions of the corresponding 
two gluino systems evaluated
at the origin. For simplicity, they have been approximated above by their 
values 
for a pure Coulombic potential, which is a good approximation for very heavy
gluino masses.  However, in our actual calculations 
we have also incorporated 
a linear confinement  term \cite{Kuhn}, which  increases 
the predicted widths by 40\% (20\%) for M=200 GeV (600 GeV). 
Although  much
larger than the decay width (\ref{i1}) of the constituent gluinos,
these decay widths are still very small compared to the gluinonium
mass $M \approx 2 m_{\tilde g}$. The size of all three states is of order 
$a^B \equiv 4(\alpha_s M)^{-1}$\/,
which is much smaller than the confinement length, thus justifying the
relative stability of  the  colour octet states(see \cite{Haber,Chikovani90}). 

To summarise, vector gluinonium 
$\psi_{\tilde{g}}^8$ is a heavy compact object which behaves rather like a
heavy gluon, except that its coupling to quarks is much stronger than its
coupling to gluons. Hence it is most readily produced via
$q \bar{q}$ annihilation, and  the Tevatron  is at least a potentially
promising place to look, being a source of  both
valence quarks and valence antiquarks. In contrast, the
pseudoscalar states $\eta_{\tilde{g}}^0$ and $\eta_{\tilde{g}}^8$  
couple predominantly to gluons, and can be produced equally well in both
$pp$ and $\bar{p}p$ collisions via the gluon-gluon fusion mechanism. 
Their production cross-section increases
more rapidly with energy than that for vector gluinonium,
and there is more chance of detecting them at
the LHC. We discuss these possibilities in turn.

\section{Vector Gluinonium at the Tevatron}

In this section  we consider 
the possibility of observing vector gluinonium as a peak in the
di-jet invariant mass distribution at the Tevatron. This possibility has
been enormously enhanced since the previous study \cite{Chikovani89},
 by improvements in the efficiency of  heavy quark jet tagging at CDF
\cite{Benlloch}  and by the increase of the Tevatron luminosity. 

The vector gluinonium is produced and decays
in $p \bar{p}$ collisions  via the subprocess
\be\label{vecprod}
q + \bar{q} \rightarrow \psi_{\tilde{g}}^8\  \rightarrow q + \bar{q}, \; Q +
\bar{Q} 
\ee
where  we use the symbols $q = u,d,s$  and $Q = c,b,t$ to distinguish
light and heavy quarks{\footnote{Obviously, $t$-quarks contribute only if
the gluino is heavy enough, and even then for the range of gluino masses
accessible at the Tevatron this contribution is strongly
suppressed by the available phase space. So, in our rough estimates we consider
just two "taggable" heavy quark flavours, though in the full simulation
correct mass-dependent branching fractions have been used.}.

The nature of the background depends on the value of the ratio
$M/\sqrt{s}$.  At  Tevatron energies, the range of interest lies 
mainly  in large values  $M/\sqrt{s}>0.2$, where the luminosity of 
colliding $q\bar{q}$ pairs prevails over the gluon-gluon pairs.
In this region, the main sources of two-jet background are the subprocesses
\begin{equation}\label{bsub}
q + \bar{q} \stackrel{QCD}{\longrightarrow} 
g + g, \;\;\; q + \bar{q},\;\;\; Q + \bar{Q} 
\end{equation}
where the first two have the angular dependence 
$$
\frac{d\sigma}{d\cos \theta^*} \propto (1 - \cos^2 \theta^*)^{-2} \; .
$$
That is, they peak sharply at $\cos \theta^* = \pm 1$, where $\theta^*$ 
denotes the scattering angle in the centre of mass frame of the two jets.
In contrast, the signal from the subprocess (\ref{vecprod})
has a much weaker dependence  
$$
\frac{d\sigma}{d\cos \theta^*} \sim 1 + \cos^2 \theta^* \; .
$$ 
Hence, a cut $|\cos \theta^*|<z$ should improve the signal-to-background ratio.
However, if the cut $z$ is too low, the signal detection efficiency is
low. The optimum value of $z$ in this case is close to $2/3$.

The usefulness of heavy quark tagging is clearly brought out by considering
the production ratios for the various final states in both the signal 
(\ref{vecprod}) and background (\ref{bsub}). The relative contribution of 
the three background subprocesses in (\ref{bsub}) 
at small $|\cos \theta^*|$ is given by \cite{Chikovani89,Eichten}
\begin{equation}
g g : q \bar{q} : Q \bar{Q} = 14 : 65 : 6,
\end{equation}
while for the signal (\ref{vecprod}) one has 
\begin{equation}
g g : q \bar{q} : Q \bar{Q} = 0 : 3 : 2.
\end{equation}
Hence  by tagging the heavy quark jets one reduces the background
by a factor of $85/6\approx14$, while retaining 40\% of the signal.
At larger $|\cos \theta^*|$ the gain in the signal-to-background ratio
caused by heavy quark jet tagging will be even larger.

At smaller gluinonium masses $M \approx 2 m_{\tilde{g}}< 200$ GeV, initial 
gluons  contribute much more significantly to the background, even with heavy 
quark jet tagging, through the subprocess
\begin{equation}\label{ggQQ}
g + g \stackrel{QCD}{\longrightarrow} Q + \bar{Q} \; .
\end{equation}
This makes the signal-to-background ratio hopelessly small for any
reasonable jet-jet invariant mass resolution. However, this region of
gluino masses is already covered by other methods.

So, most of the two-jet QCD background at large invariant masses arises 
from  light quark and gluon jets, and the
signal-to-background ratio can be significantly enhanced by 
triggering on heavy quark jets \cite{Chikovani89}.
To check that this makes the detection of vector gluinonium a viable 
possibility at the Tevatron, 
we have simulated both the gluinonium signal and the 2-jet QCD background
using the PYTHIA 5.7   Monte Carlo event generator \cite{PYTHIA}.
The vector gluinonium production and decay was simulated
by exploiting the fact that
$\psi_{\tilde{g}}^8$ behaves much like a heavy $Z^{\prime}$ with axial current
and lepton couplings set to zero and a known mass-dependent vector current
coupling to quarks, chosen to comply with the corresponding decay width
after
taking into account appropriate colour and flavour counting. This effective
coupling included the non-Coulomb corrections mentioned earlier, 
and an enhancement due to the fact that numerous radial excitations of the
$\psi_{\tilde{g}}^8$, which could not be separated from it for any
reasonable mass resolution, will also contribute. These yield an overall  
factor 
of between 1.8 and 1.6 depending on $M$, and the resulting  effective vector 
coupling $a_V$ falls exponentially from $a_V=0.225$ at 
$M=2m_{\tilde g}=225$ GeV to $a_V=0.172$ at $M=2m_{\tilde g}=450$ GeV. 
This signal sits on a much larger background, which has been simulated on 
the assumption
that it arises entirely from the leading order QCD subprocesses for
heavy quark pair production (\ref{bsub}) and (\ref{ggQQ}). A constant
$K$ factor $K=2.0$ has been used for both signal and background.

The cross section for vector gluinonium production at
the Tevatron is shown in Fig. 3 for $\sqrt{s}=1.8$ TeV. Only decays into heavy 
quark-antiquark pairs were taken into account, and a heavy quark jet
tagging efficiency of $50\%$ was assumed. The cut on the jet angle $\theta^*$
in the two-jet c.m.s. frame is $|\cos \theta^*|<2/3$, and the cut on 
jet rapidity $y$ is $|y|<2.0$. The analysis shows that
with an integrated luminosity of 200 pb$^{-1}$ one can hope to see
the gluinonium signal from
gluinos with masses up to 140 GeV as a 5 standard deviation peak,
and the signal from gluinos with masses up to 170 GeV as a 3 standard 
deviations peak. The signal-to-background
ratio is around $7-10\%$ at the peak for the assumed two-jet invariant mass
resolutions of 25 GeV, 30 GeV and 38 GeV at
$M=225$ GeV, 320 GeV and 450 GeV respectively {\cite{Benlloch,cdfj}}.

For an upgraded Tevatron with the energy increased to 2.0 TeV and an 
integrated
luminosity of 1000 pb$^{-1}$, the reach is significantly higher.
The cross-section obtained with the same cuts and the same assumptions about 
the detector resolution
and heavy quark jet tagging efficiency is shown in Fig. 4. In this case
one can hope to see
the gluinonium signal from
gluinos with masses up to 220 GeV as a 5 standard deviation peak,
and the signal from gluinos with masses up to 260 GeV as a 3 standard 
deviation peak.
Finally, we note that the statistical  significance of the peak is essentially
inversely proportional to the two-jet invariant mass resolution, so the
reach can be significantly extended if some way is found to improve the latter.

\section{Pseudoscalar Gluinonium at the LHC}

Pseudoscalar gluinonium can also be detected as a
peak in the the di-jet invariant mass distribution. However, both 
$\eta_{\tilde{g}}^0$ and $\eta_{\tilde{g}}^8$ decay predominantly into 
gluon jets, which are also the main component of the background. Quark jet 
tagging does not help in this case, and one must rely on high statistics and 
good resolution
to separate the signal from the background. These can hopefully be achieved 
at LHC, as we shall now discuss.

Pseudoscalar gluinonium is produced, as we have noted, via the subprocess
\be\label{etaprod}
g + g \rightarrow  \eta_{\tilde{g}}^{0,8}  \rightarrow g + g 
\label{prod}
\ee
while the 2-jet background in $pp$ collisions is 
dominated by the gluon scattering QCD subprocess
\be\label{etabk}
g + g \stackrel{QCD}{\longrightarrow} g + g
\ee
with some contribution from the  valence light quark elastic
scattering subprocesses  
\be\label{etabk1}
g + q \stackrel{QCD}{\longrightarrow} g + q,\;\;\;
q + q \stackrel{QCD}{\longrightarrow} q + q
\ee
at higher values of $M/\sqrt{s}$.
Once again, the angular dependence of the background  has 
large peaks in the forward and backward directions,
$d\sigma/d\cos \theta^*\propto (1 - \cos^2 \theta^*)^{-2}$, while the
signal is isotropic in the two jet centre of mass frame. Thus a cut 
$|\cos \theta^*|<z$ with optimal $z$ close to 0.6 will increase the
signal-to-background ratio.

To check that  pseudoscalar gluinonium detection is a viable 
possibility at the LHC, we have
again simulated the QCD background, including all the above subprocesses, 
using the PYTHIA 5.7 generator \cite{PYTHIA}. Gluinonium production has not 
been explicitly implemented in PYTHIA, but
all we need is a pseudoscalar object of the right mass which is produced by
gluon gluon fusion and decays to two gluons. In other words, pseudoscalar
gluinonium  behaves like a
pseudoscalar state of heavy quarkonia with  appropriately adjusted mass,
couplings and width. In this way the signal can also be simulated in PYTHIA,
where we have again used a K-factor of 2.0 and an
enhancement factor of 1.5 to allow for the fact that numerous radial 
excitations of the  $\eta_{\tilde{g}}^0$ and 
$\eta_{\tilde{g}}^8$, which could not be separated from them for any
reasonable mass resolution, will also contribute. In addition one can not
separate the
signals from $\eta_{\tilde{g}}^0$ and $\eta_{\tilde{g}}^8$ either, so a 
single effective width $\Gamma$ has been used
to describe
the production of all pseudoscalar gluinonium states. This effective 
width was calculated by summing properly weighted widths (\ref{geta0}) and
(\ref{geta8}), with non-Coulomb corrections and higher resonance contributions
included, giving $\Gamma =$ 0.85 GeV, 1.1 GeV and 1.6 Gev for $M=$200 GeV, 
600 GeV and 2 TeV, respectively.

The resulting cross section for pseudoscalar gluinonium production at
the LHC is shown in Fig. 5. The cut on the jet angle $\theta^*$
in the two-jet centre of mass frame is $|\cos \theta^*|<2/3$, and the cut on 
jet rapidity $y$ is $|y|<2.5$. The estimated
two-jet invariant mass resolution varies from 22 GeV at $M=300$ GeV
to 32 GeV at $M=600$ GeV and 50 Gev at $M=2$ Tev \cite{Chikovani93}.
The signal-to-background 
ratio is smaller than in the previous section, of order 1-2$\%$, 
but the much higher luminosity of LHC leads none the less
to a higher reach in gluino mass. As can be seen in Fig. 5, with a
luminosity of $10^4$ pb$^{-1}$ one expects to reach the gluino masses
800 (1000) GeV at 5 (3) standard deviations, while for the highest likely
integrated luminosity of $2\cdot 10^5$ pb$^{-1}$ gluino masses up to
1300 (1500) GeV could be covered.

\section{Conclusions}

We have studied the production characteristics of various bound states of 
two gluinos (gluinonium) for existing and future hadronic colliders. We
conclude that gluinonium states can be detected as narrow peaks in the
di-jet invariant mass spectra, effectively complementing more traditional
gluino searches, in the case when the gluino is lighter than the
squarks. 

In $p\bar p$ collisions one expects copious production of 
vector gluinonium, which decays predominantly to $q\bar q$ pairs.
The high efficiency of the heavy quark jet tagging at  CDF,
together with the boost of the Tevatron luminosity, should allow one to reach
gluino masses of 140-170 GeV at $\sqrt{s}=1.8$ TeV, and 200 pb$^{-1}$
and 220-260 GeV at $\sqrt{s}=2.0$ TeV and 1000 pb$^{-1}$, with
realistic efficiencies, resolutions and experimental cuts taken into
account.
 
In $pp$ collisions pseudoscalar gluinonium production is quite large,
again  allowing one to look for a narrow peak in di-jet invariant mass spectra
in spite of the high background from QCD scattering subprocesses. The high
luminosity of the LHC, together with the high design resolutions of the ATLAS 
and CMS detectors, should allow one to reach gluino
masses of $800-1000$ GeV at $\sqrt{s}=14$ TeV and $10^4$ pb$^{-1}$,
while for the highest likely integrated luminosity  $2\cdot10^5$ pb$^{-1}$
the reach increases to $1300-1500$ GeV.

Finally, we note that if gluinonium is observed, its mass can be determined 
with an accuracy of a few GeV. This makes gluinonium detection by far the most
precise way of measuring the gluino mass, since the binding energy inside 
gluinonium is very small and calculable.

\newpage
\begin{center}

\bigskip

\vspace{0.5cm}

\begin{tabular}{llll}
&&& \\
\cline{1-4}
&&& \\
    & $ 1 $ & $ 8_S$ & $ 8_A $  \\
&&& \\
\cline{1-4}
&&& \\
$ ^1S_0$ & $0^- \;  (\eta_{\tilde{g}}^0)$ &  $0^-  \; 
(\eta_{\tilde{g}}^8)$ &   \\
$ ^3S_1$ &  &  & $1^- \; (\psi_{\tilde{g}}^8)$   \\
&&& \\
$ ^1P_1$ &  &  & $1^+ $   \\
$ ^3P_0$ & $0^+ $ &  $0^+ $ &   \\
$ ^3P_1$ & $1^+ $ &  $1^+ $ &   \\
$ ^3P_2$ & $2^+ $ &  $2^+ $ &   \\
&&& \\
$ ^1D_2$ &  &  & $2^- $   \\
$ ^3D_1$ & $1^- $ &  $1^- $ &   \\
$ ^3D_2$ & $2^- $ &  $2^- $ &   \\
$ ^3D_3$ & $3^- $ &  $3^- $ &   \\
&&& \\
\cline{1-4}

\end{tabular}

\end{center}

\vspace{0.5cm}

\noindent
{\bf Table 1.} Spin-parities $J^P$ for the low lying states of gluinonium 
with L $\le$ 2. The three
columns correspond to the colour singlet state 1 and the  
symmetric and antisymmetric colour octet states $8_S$ and $8_A$ respectively.


      \newpage                                                       
        \begin{figure}[top]
        \begin{center}
        \mbox{
        \epsfysize=6cm
        \epsfbox{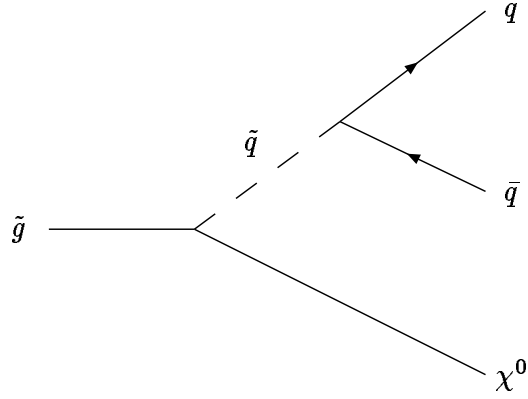}
        }
        \end{center}            
        \caption[1]{ The dominant decay  mode for gluinos, if they are lighter
        than the squarks, where $\chi^0$ is a neutralino. }
        \end{figure}

        \begin{figure}[top]
        \begin{center}
        \mbox{
        \epsfysize=3.3cm
        \epsfbox{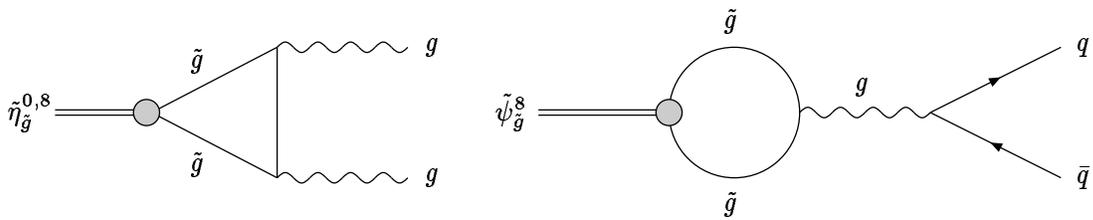}
        }
        \end{center}            
        \caption[2]{ The dominant decay mechanisms for (a) pseudoscalar and
        (b) vector gluionium. The dominant production processes in hadron
        collisions are obtained by reading the diagrams from the right. }
        \end{figure}            
        
       \newpage

        \begin{figure}[top]
        \begin{center}
        \mbox{
        \epsfysize=15cm
        \epsfbox{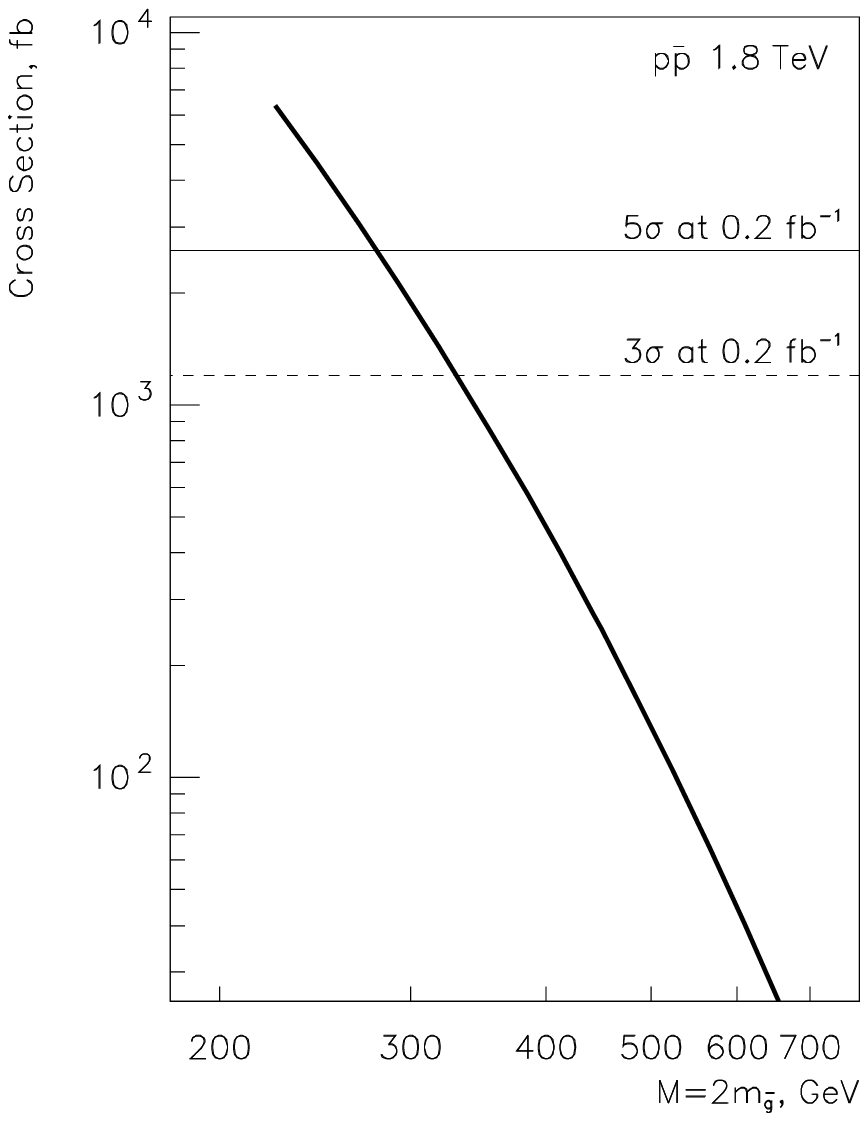}
        }
        \end{center}            
        \caption[3]{The calculated  production cross section of
         vector gluinonium in $p\bar p$ collisions at  1.8 TeV. The solid and 
         broken horizontal lines indicate the cross sections corresponding to 
         a statistical significance at the peak of 5 and 3 standard deviations 
         respectively, for a luminosity of 200 pb$^{-1}$. (See the text for 
         the cuts and resolutions used).}
        \end{figure}            

       \newpage
                                                       
        \begin{figure}[top]
        \begin{center}
        \mbox{
        \epsfysize=15cm
        \epsfbox{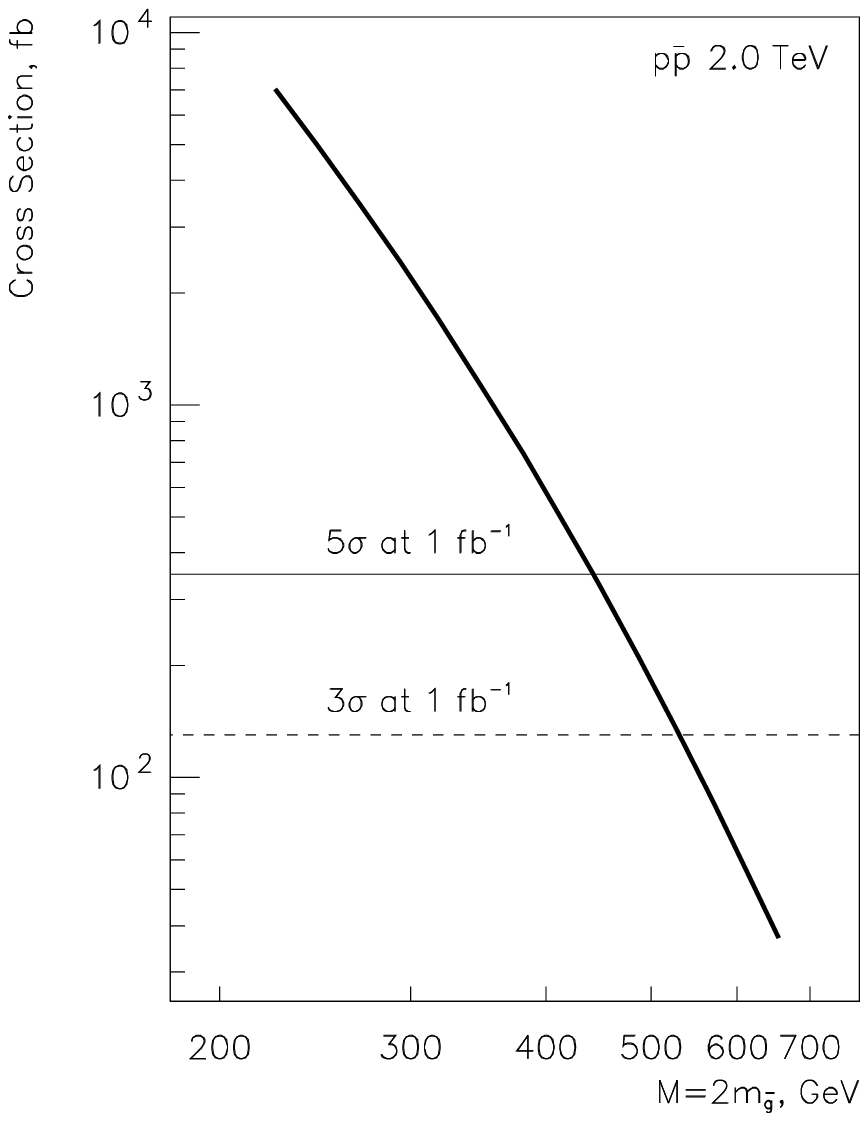}
        }
        \end{center}            
        \caption[4]{ The calculated  production cross section of
        vector gluinonium in $p\bar p$ collisions at 2.0 TeV. The solid and 
        broken horizontal lines indicate the cross sections corresponding to a 
        statistical significance at the peak of 5 and 3 standard deviations 
        respectively, for a luminosity of 1 fb$^{-1}$. (See the text for the 
        cuts and resolutions used).}
        \end{figure}            

        \newpage

        \begin{figure}[top]
        \begin{center}
        \mbox{
        \epsfysize=15cm
        \epsfbox{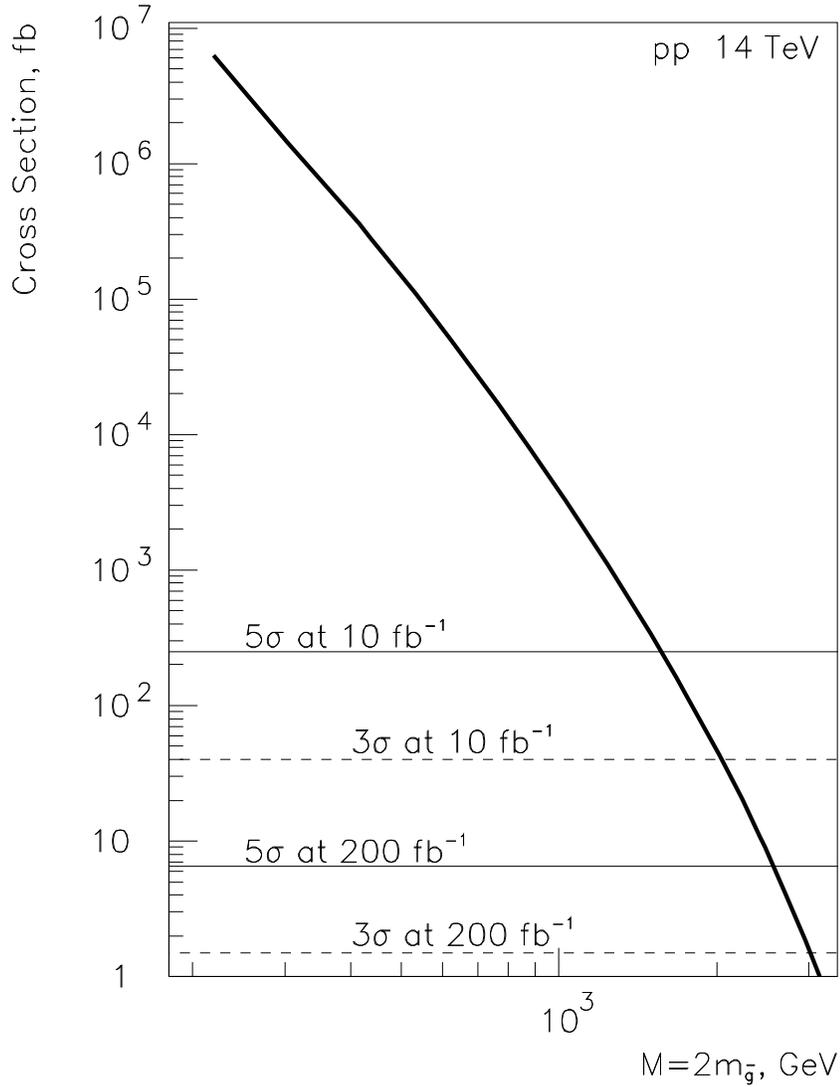}
        }
        \end{center}            
        \caption[5]{The calculated  production cross section of
pseudoscalar gluinonium in  $pp$ interactions at 14 TeV. 
The solid and broken
horizontal lines indicate the cross sections corresponding to a 
statistical significance at the peak of 5 and 3 standard deviations 
respectively, for two
values of the  luminosity: $10^4$ pb$^{-1}$ and 
$2\cdot10^4$ pb$^{-1}$. (See the text for the cuts and 
resolutions used).}
         \end{figure}

\end{document}